\documentclass[aps, pra, reprint, superscriptaddress]{revtex4-1}

\usepackage{amsfonts}
\usepackage{amsmath}
\usepackage{amssymb}

\usepackage{array}
\usepackage{dcolumn}

\usepackage{hyperref}
\hypersetup{colorlinks = true,
  linkcolor = blue,
  citecolor = blue,
  urlcolor = blue,
  filecolor = blue}

\usepackage{graphicx}

\begin{document}

\title{Bosonic Szilard Engine Assisted by Feshbach Resonances}
\author{J. Bengtsson}
\author{M. Nilsson Tengstrand}
\author{S.M. Reimann}
\email{reimann@matfys.lth.se}

\affiliation{Mathematical Physics and NanoLund, Lund University, Box 118, 22100
  Lund, Sweden}

\date{\today}

\begin{abstract}
  It was recently found that the information-to-work conversion in a quantum Szilard
  engine can be increased by using a working medium of bosons with attractive
  interactions. In the original scheme, the work output depends on the insertion and
  removal position of an impenetrable barrier that acts like a piston, separating
  the chambers of the engine. Here, we show that the barrier removal process can be
  made fully reversible, resulting in a full information-to-work conversion if we
  also allow for the interaction strength to change during the cycle. Hence, it
  becomes possible to reach the maximum work output per cycle dictated by the second
  law of thermodynamics. These findings can, for instance, be experimentally
  verified with ultra-cold atoms as a working medium, where a change of interaction
  strength can be controlled by Feshbach resonances.
\end{abstract}

\maketitle

\section{Introduction}
Szilard's famous single-particle engine~\cite{szilard1929} is the archetype of an
information heat engine that extracts work through measurement and feedback
operations. The original setup consists of a single particle in a container that is
coupled to a single heat bath. Dividing the container into two equally sized parts
by a movable piston,  the engine employs a type of ``Maxwell's demon'' that provides
the information on which side of the piston the particle resides (measurement). Work
may then be extracted from the particle's collisions with the piston that pushes it
aside (feedback). The connection between information and work is ensured by the
second law of
thermodynamics~\cite{Landauer1961,Bennett1982,Bennett2003,Leff2003,Plenio2001,Sagawa2008,
  Sagawa2009,parrondo2015}:  Erasing one bit of information costs at least the entropy
$k_B \ln 2$, where $k_B$ is the Boltzmann constant.
Szilard's thought experiment dates back almost a century and has led to a plethora
of studies,  for example  addressing the physicality of the measurement and erasure
processes~\cite{Landauer1961,Bennett1982,Bennett2003}, or investigating the role of
information in thermodynamics in general~\cite{parrondo2015}.

A highly interesting question is, how thermodynamic properties are changed when quantum effects are taken into account.
Different quantum versions of Szilard's engine have been suggested, with
single-particle~\cite{Zurek1986} to many-body working
media~\cite{kim2011,kim2012,Cai2012,Lu2012,Zhuang2014,Plesch2014,Jeon2016,bengtsson2018}.
For engines with non-interacting particles, bosons were found superior to
fermions~\cite{kim2011}.
As first shown for two particles~\cite{kim2012} and recently generalized to the
many-body regime~\cite{bengtsson2018}, attractive interactions between the bosons
can enhance the information-to-work conversion even further.

Quantum effects in correlated many-particle
systems may  enhance the performance also of other kinds of quantum heat
engines. For instance, it was recently
found~\cite{Jaramillo2016} that a quantum Otto engine~\cite{Quan2007} with a
many-particle working medium consisting of an interacting Bose
gas, confined in a time-dependent harmonic trap, is able to outperform a
corresponding ensemble of single-particle quantum heat engines.

Szilard-like information-to-work conversion has been experimentally demonstrated
for classical systems~\cite{toyabe2010, roldan2014, koski2014,koski2015}, 
but the corresponding setups in the quantum
realm~\cite{Zurek1986,kim2011,kim2012,Plesch2014,bengtsson2018} have so far evaded
realization. However, as noted previously in the literature, see e.g.
Refs.~\cite{kim2011,bengtsson2018}, such experiments could indeed become possible
with ultra-cold atoms. With nowadays quantum-optical trapping techniques, the shape and
dimensionality of the confining potential can be modified with a very high degree of
experimental control~\cite{Bloch2008}, bringing the realization of the quantum
Szilard engine closer to reality.
\begin{figure}
  \includegraphics[width = 0.3\textwidth]{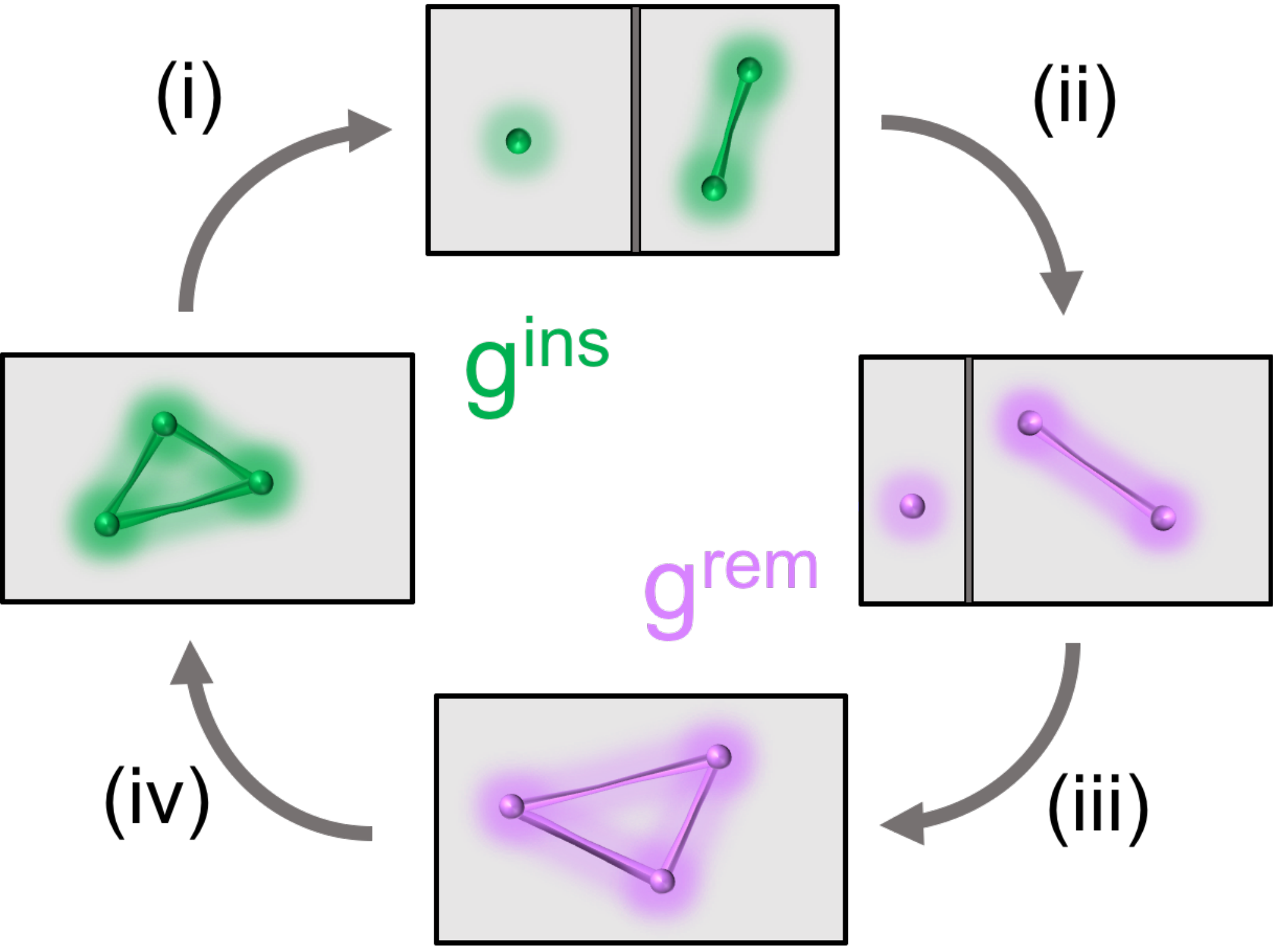}
  \caption{Schematic picture of a Szilard engine cycle where
    both the removal position of the barrier and the strength of interactions at
    barrier insertion and barrier removal, $g^{\mathrm{ins}}$ and
    $g^{\mathrm{rem}}_n$, can be varied. In this particular cycle, one particle was
    found to the left (and two to the right) of the barrier after insertion.}
  \label{fig:cycle}
\end{figure}

A moving barrier as in the case of the Szilard engine, or a confinement chamber that
changes its size as in the Otto cycle, are not the only ways to produce work.
Recently it was suggested~\cite{li2017} that a change in the interaction strength,
by means of Feshbach resonances~\cite{chin2010}, can be used to extract work in a
bosonic quantum Otto engine. Here, we investigate how a similar concept applies to
the bosonic quantum Szilard engine. How does a Feshbach-guided Szilard cycle compare
with Szilard's original concept of a moving piston? As a first step, we keep the
position of the impenetrable barrier fixed and vary only the strength of the
interactions during the cycle. We then allow for both  the interaction strength and
the barrier position to vary during the cycle.

To quantify the performance of the many-particle engine, we determine its average
work output
relative to the value of $k_BT\ln 2$ obtained for the original (single-particle)
Szilard engine connected to a heat bath at temperature $T$. We find that for the
Feshbach-driven engine, i.e. the engine with fixed barrier position, the maximum
relative work output  oscillates with the number of bosons, and for larger $N$
indeed exceeds the corresponding Szilard engine with non-interacting bosons and a
movable barrier~\cite{kim2011,kim2012}. Yet, the maximum relative work output,
obtained by changing the interaction alone, is smaller than that of an engine with a
movable barrier and constant attractive interactions between the
bosons~\cite{bengtsson2018}.

The average work output can be significantly increased if we combine
the possibility to tune the interaction strength with the usual
Szilard engine setup with a movable wall. In fact, in the deep quantum
regime, we find that it is possible to construct a Feshbach-assisted
protocol that maximize both the information and the
information-to-work conversion efficiency for an arbitrary number of
bosons. In other words, by varying the barrier position together with
the interaction strength, it is possible to construct an engine that
produces the maximum possible work output in the low-temperature
limit. The maximal work output is here encountered for working media
that undergo a transition between a non-interacting Bose gas and a
Tonks-Girardeau gas~\cite{Girardeau1960}. 

The paper is organized as follows: In Sec.~II we describe the general quantum
Szilard cycle, where both the interactions between the particles and the position of
the barrier may be varied in order to maximize the work output. In Sec.~III we
discuss the work output
for the case of a fixed barrier position at the center of the container, where the
engine is driven by the variation of the interaction strength alone. Sec.~IV then
discusses the work output obtained with a simultaneous variation of the interaction
strength and the barrier position, i.e. when the conventional Szilard engine is
assisted by Feshbach resonances. Finally, in Sec.~V, we conclude with  prospects for future work.

\section{Szilard Cycle for Bosons with variable interactions}
For the setup of the Szilard engine, we consider a one-dimensional infinite well of
length $L$, i.e. a hard-wall box potential with $V(0)=V(L)\rightarrow \infty $ and
$V(x)=0$ for $0<x<L$. The trap confines a small number $N$ of spinless bosons
interacting by the pseudo-potential of contact type, $g\delta(x_1 - x_2)$, as
commonly used for ultra-cold atomic gases~\cite{Bloch2008}. Here, $g$ is the
strength of the two-body interaction, given in
units of $\tilde g = \hbar ^2/(mL)$.

The steps of the Feshbach-assisted quantum many-particle  Szilard cycle are
illustrated in Fig.~\ref{fig:cycle}: (i) The working medium of $N$ bosonic particles
confined in the box, interacting with some initial interaction strength
$g=g^{\mathrm{ins}}$, is split into two parts by an impenetrable barrier at a
position $x=\ell^\mathrm{ins}$; (ii) after separation, the number $n$ of particles
on, say,  the left side of the wall is measured,  and  the barrier is then moved to
the position $\ell_n^\mathrm{rem}$, expanding one side of the chamber and
compressing the other. In addition, according to the measurement outcome, the
interaction strength is changed to $g_n^{\mathrm{rem}}$. In step (iii) the barrier
is removed,  and finally the interaction is tuned back to its initial value
$g^{\mathrm{ins}}$ in step (iv).

We assume that all processes are carried out quasi-statically and isothermally in
contact with a single heat bath. At a given temperature $T$,
the change in the partition function
\begin{equation}
  Z = \sum _j \hbox{e}^{-E_j/(k_BT)}
  \label{partition_function}
\end{equation}
determines the work output associated with the isothermal process,
\begin{equation}
  W\le -\Delta F = k_B T \Delta (\hbox{ln} Z),
  \label{work_free}
\end{equation}
where $F$ is the Helmholtz free energy. In Eq.~(\ref{work_free}), equality is
reserved for reversible processes. The sum in Eq.~(\ref{partition_function}) runs
over the full spectrum of $N$-body eigenenergies $E_j$, which, in turn, depend both
on the interaction between the particles and the position of the barrier. For the
one-dimensional system of bosons with contact interaction, we use the Bethe
ansatz~\cite{gaudin1971} to find these energies. For instance, the contribution to
$E_j$ from the $n_j$, where $0 \leq n_j \leq N$, bosons located to the left of the
barrier at $\ell$ is here given by $(\hbar ^2/2m)\sum_{\alpha=1}^{n_j} k_{(j,\alpha
  )}^2$,  where the $k_{(j, \alpha )}$ are the solutions to a set of coupled
transcendental equations
\begin{align}
  \ell k_{(j,\alpha )} =& \pi b_{(j,\alpha )} + \sum_{\beta \neq \alpha }
  \tan^{-1}\left(\frac{gm}{\hbar^2}\frac{1}{k_{(j,\alpha )} - k_{(j,\beta )}} \right)  \nonumber\\
  &+\sum_{\beta \neq \alpha } \tan^{-1}\left(\frac{gm}{\hbar^2}\frac{1}{k_{(j, \alpha )} +
    k_{(j,\beta )}} \right),
  \label{Eq:Bethe}
\end{align}
The integers $b_{j,\alpha}$ are ordered as $1 \leq
b_{(j,1)} \leq b_{(j,2)} \leq \cdots \leq b_{(j,n_j)}$.
This system of equations can be recast as a non-linear least squares problem,
which we address using a standard trust region algorithm. For $g<0$, the solutions
$k_{(j,\alpha)}$ can be complex-valued, which may complicate the numerics (and which
effectively limits the negative values of $g^\mathrm{ins}$ considered in this work).

The average work output of the full cycle (not accounting for memory processing) of
the Feshbach-guided Szilard engine is given by (see Appendix)
\begin{equation}\label{workfa}
  W = -k_B T \sum_{n = 0}^{N} p_n(\ell^{\mathrm{ins}}, g^{\mathrm{ins}})\ln \left[
    \frac{p_n(\ell^{\mathrm{ins}}, g^{\mathrm{ins}})}{p_n(\ell^{\mathrm{rem}}_n,
      g_n^{\mathrm{rem}})} \right],
\end{equation}
where $p_n(\ell, g)$ is the probability to find $n$ particles to the left of a
barrier inserted at $\ell$ when the interaction strength is $g$. All processes
except the barrier removal are assumed reversible. It should also be noted that for
the isothermal processes considered here, the order by which the expansion and
interaction tuning steps (ii) are carried out does not matter. This is due to the
fact that for reversible isothermal processes, only the initial and final states
determine the work of the processes (see Eq.~(\ref{work_free})). The probabilities
$p_n(\ell^{\mathrm{rem}}_n, g_n^{\mathrm{rem}})$ in Eq.~(\ref{workfa}) characterize
the reversibility of the engine \cite{Jeon2016}, and can be thought of as the
probabilities to return to a certain configuration if the removal process is
performed in reverse. If all $p_n(\ell^{\mathrm{rem}}_n, g_n^{\mathrm{rem}}) = 1$,
also the removal process is made reversible and Eq.~(\ref{workfa}) reduces to
$W=k_BTI$, where $I=-\sum_{n=0}^{N}p_n(\ell^{\mathrm{ins}},
g^{\mathrm{ins}})\ln[p_n(\ell^{\mathrm{ins}}, g^{\mathrm{ins}})]$ is the Shannon
information. This information is maximized by a uniform probability distribution,
and the work output of the Szilard engine with $N$ quantum particles is consequently
bounded according to $W \leq k_B T\ln(N+1)$.

\section{Feshbach-Driven Szilard Engine}
Let us first consider a bosonic Szilard engine that is driven by a change of the
interaction strength alone, with insertion and removal of the barrier at the central
position, $\ell^{\mathrm{ins}} = \ell_n^{\mathrm{rem}} = L/2$. This particular
choice of barrier position maximizes the Shannon information for non-interacting
bosons in the $T\rightarrow 0$ limit.
The change in interaction strength from $g^{\mathrm{ins}}$ before insertion to
$g_n^{\mathrm{rem}}$ at removal of the barrier now plays a role similar to the
change in barrier position in the conventional Szilard cycle.
For example, let us consider the two-particle engine and cycles in which one boson
is measured on either side of the barrier. Simply removing the barrier without any
change in $g$, i.e. with $g^{\mathrm{rem}}_1 = g^{\mathrm{ins}}$, the average work
gain equals the average work cost of introducing the barrier in the first place 
and  no net work output is thus possible. 
If we instead first increase $g$ (which costs no
work when only one particle is on either side) such that $g^{\mathrm{rem}}_1 >
g^{\mathrm{ins}}$, the amount of work that can be extracted from removing the
barrier is reduced. However, once the barrier is removed, we may extract an
additional amount of work by decreasing the interaction strength to its initial
value of $g^{\mathrm{ins}}$. In total, the combined work gain is now larger than the
cost of introducing the barrier. The positive net work output can, in this case, be
explained by the fact that also the 
losses caused by tunneling particles during the barrier removal process, are
reduced when $g^{\mathrm{rem}}_1$ is increased.

\begin{figure}
  \centering
  \includegraphics[width = 0.5\textwidth]{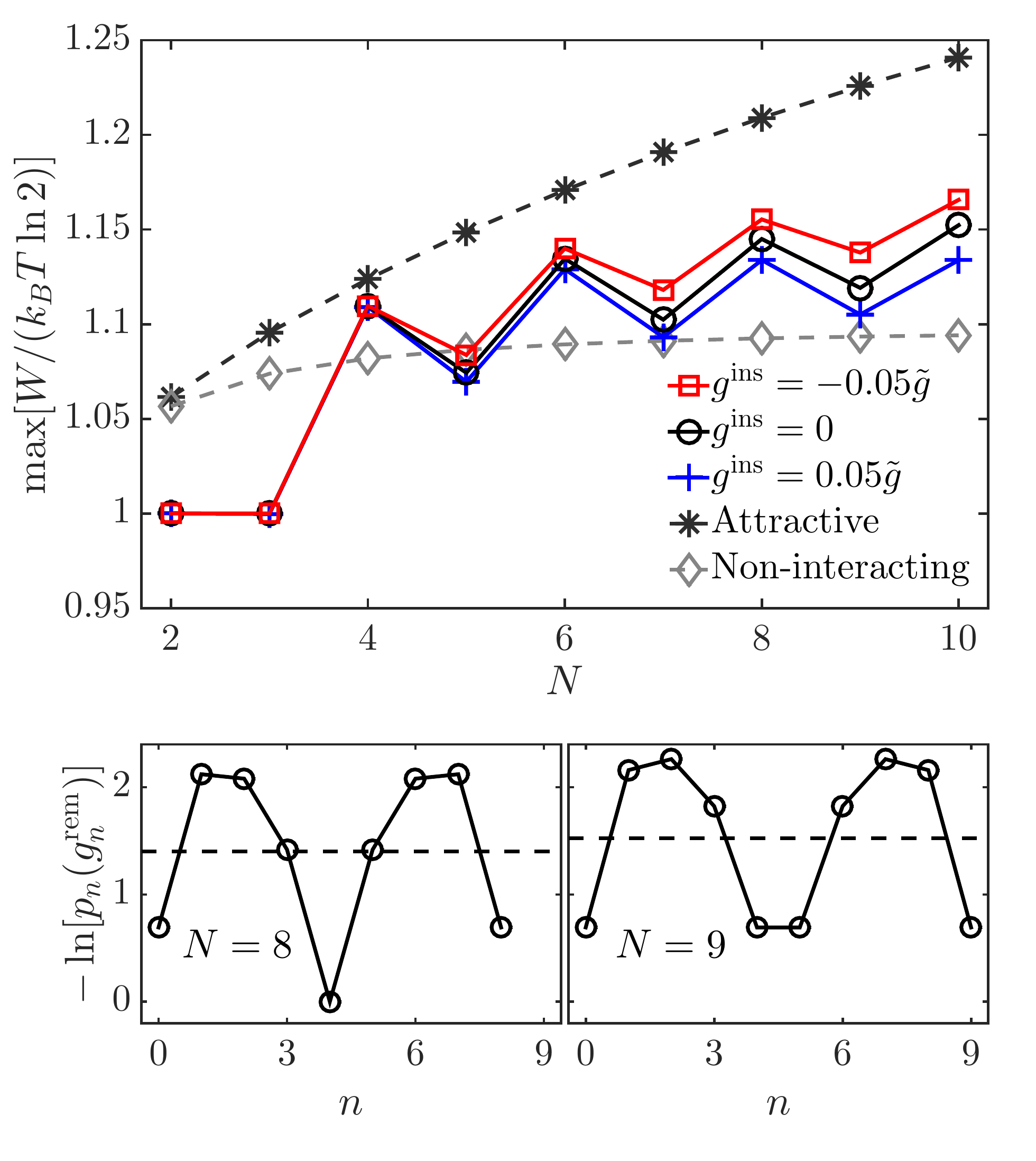}
  \caption{(\textit {Color online}) Maximum of the ratio between the
    average work output $W$ of the Feshbach-driven Szilard engine and
    the value $k_BT\ln 2 $, shown for different number of bosons, $N$. The
    different interaction strengths during the removal of the barrier
    are optimized to maximize $W$ for each considered $T$.
    Three different
    interaction strengths at barrier insertion are considered, namely
    $g^{\mathrm{ins}}=-0.05\tilde{g}$ (\textit{red line with squares}),
    $g^{\mathrm{ins}}=0$ (\textit{black line with circles}), and
    $g^{\mathrm{ins}}=0.05\tilde{g}$ (\textit{blue line with crosses}), where
    $\tilde{g} = \hbar^2/(mL)$. For reference we also include the
    corresponding data for the conventional $N$-particle Szilard
    engine with a movable barrier.
    The work output of non-interacting bosons (\textit{dashed light
      grey line with diamonds)} is seen together with that of bosons with weak constant
    attraction (\textit{dashed dark grey line with stars}). \textit{Lower panels:}
    The minimum losses
    $-\ln[p_n(L/2,g^\mathrm{rem}_n)]$ in the information-to-work conversion for
    $N=8$ and $N=9$ bosons with $g^{\mathrm{ins}}=0$ and at $k_B T =
    \varepsilon_1$, where $\varepsilon_1 = \hbar^2 \pi^2/(2mL^2)$ is the
    single-particle ground state energy in the absence of a barrier.
    The black dashed lines show the average conversion losses,
    $-1/(N+1)\ln[p_n(L/2,g^\mathrm{rem}_n)]$.  }
  \label{Fig:2}
\end{figure}

Depending on the measurement outcome, the values of
$g_n^{\mathrm{rem}}$ are here chosen from a numerical sweep such that
they maximize the relative work output $W/(k_BT\ln 2)$.  These maxima
are shown in Fig.~\ref{Fig:2} for different number of bosons and for
different choices of $g^{\mathrm{ins}}$. With the exception of $N=2$,
the maximal values of $W/(k_BT\ln 2)$ are encountered in the low temperature limit for
$g^{\mathrm{ins}} =0$ and at finite temperatures for (finite) $g^{\mathrm{ins}} \neq
0$.
The strong odd/even oscillatory behavior in $N$ is
in sharp contrast to that of the ordinary Szilard cycle (here shown
for reference as grey dashed lines in Fig.~\ref{Fig:2}, both for the
non-interacting~\cite{kim2011} and the weakly
interacting~\cite{bengtsson2018} case), which shows a smooth increase
in $W/(k_BT\ln 2)$ with $N$.

To understand the origin to the oscillatory $N$-behavior seen in Fig.~\ref{Fig:2}, a closer examination of the probabilities $p_n(L/2, g^\mathrm{rem}_n)$ is called for.
The two probabilities $p_0(L/2, g^\mathrm{rem}_0)$ and $p_N(L/2, g^\mathrm{rem}_N)$
are maximized in the limit of a strong attractive interaction
\begin{equation}
  p_0(L/2, g^\mathrm{rem}_0 \to -\infty) = p_N(L/2, g^\mathrm{rem}_N\to -\infty) \to
  1/2.\label{Eq:probabilities0}
\end{equation}
Contrary, the limit of a strong repulsive interaction maximizes the probability for a
configuration with an equal, or almost equal, number of particles on either side of the barrier,

\begin{align}
  p_{N/2} ({L/2}, g^{\mathrm{rem}}_{N/2}\to \infty ) &\to 1,&&\mathrm{for~even~} N,
  \label{Eq:probabilities1} \\
  p_{(N\pm 1)/2} (L/2, g^\mathrm{rem}_{(N\pm 1)/2}\to \infty )  &\to 1/2,&&\mathrm{for
    ~odd~}N.
  \label{Eq:probabilities2}
\end{align}
The central features in the optimal probability distribution, associated with the
barrier removal, are thus slightly different for an even and for an odd number of
particles, with a fully reversible removal process possible for the former. 
In Fig.~\ref{Fig:2}, we show the corresponding minimum losses, $-\ln p_n(L/2,
g^{\mathrm{rem}}_n)$, associated with the information-to-work conversion and the
average value $-1/(N+1)\sum_n p_n(L/2, g^{\mathrm{rem}}_n)$ for  $N=8$ and $N=9$
bosons  at $k_B T=\varepsilon_1$ (which is sufficiently low for $p_n(L/2, 0)$ to be
approximately uniformly distributed) when $g^\mathrm{ins}=0$. Here $\varepsilon _i
=\hbar ^2\pi^2 i^2/(2mL^2)$ is the single-particle energy in the absence of a
barrier. We see that the average conversion loss 
is higher  for $N=9$ than for $N=8$. 
Even though the information grows like $\ln (N+1)$ in the low-temperature-limit, the increase in conversion losses, going from an even number $N$ of particles to $N+1$, turns out to be large enough for the total work to decrease. If we instead add two particles, keeping the number of particles odd or even, the increase in information dominates over the increase in losses. The general overall trend is thus that the maximum in $W/(k_BT\ln 2)$ increases with $N$, but with an oscillatory odd/even modification.

\begin{figure}
  \centering
  \includegraphics[width = 0.5\textwidth]{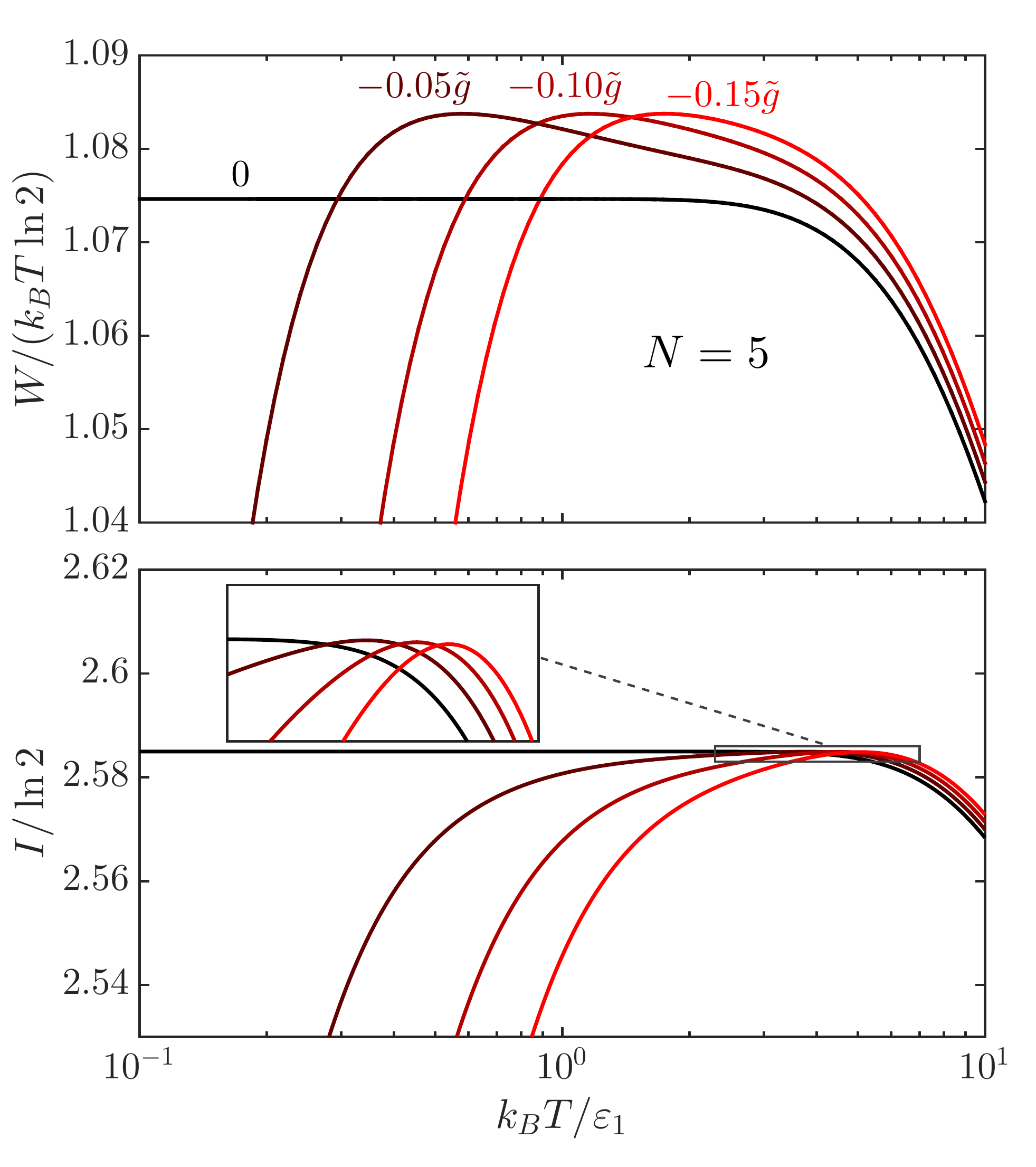}
  \caption{ (\textit{Color online}) \textit{Upper panel:} Work output of the
    Feshbach-driven Szilard engine for $N=5$ as a function of temperature and for
    different interaction strength $g^{\mathrm{ins}}$.  \textit{Lower panel:}
    Information $I = -\sum_n p_n(L/2,g^{\mathrm{ins}})$ as a function of temperature
    for the same systems. For a fully reversible cycle, $W=k_BTI$. 
  }
  \label{fig:wtdep}
\end{figure}

For $N \geq 4$, we also observe (see Fig.~\ref{Fig:2}) that a higher work ratio,
$W/(k_BT\ln2)$, may be achieved for bosons with attractive interactions, compared to
non-interacting ones, during the barrier insertion. In general, in the search for
the maximize in $W/(k_B T)$ it is important to account for the losses in the
information-to-work conversion. The maximum in $W/(k_B T)$ is thus not necessarily
seen at the maximum Shannon information. In other words, the fact that $I$ is
maximal for non-interacting bosons in the zero temperature limit does not guarantee
that also $W/(k_B T)$ is the highest possible with this particular setup. For attractive bosons, the higher
values of $W/(k_BT) $ are instead obtained at finite temperatures where both the
information-to-work conversion efficiency and the Shannon information are relatively
high (but neither of them are maximal), see Fig.~\ref{fig:wtdep} for an engine with
$N=5$. Similar features have also been seen in the conventional $N$-particle Szilard
engine with a movable barrier, as further discussed in Ref.~\cite{bengtsson2018}.

\begin{figure}
  \centering
  \includegraphics[width = 0.5\textwidth]{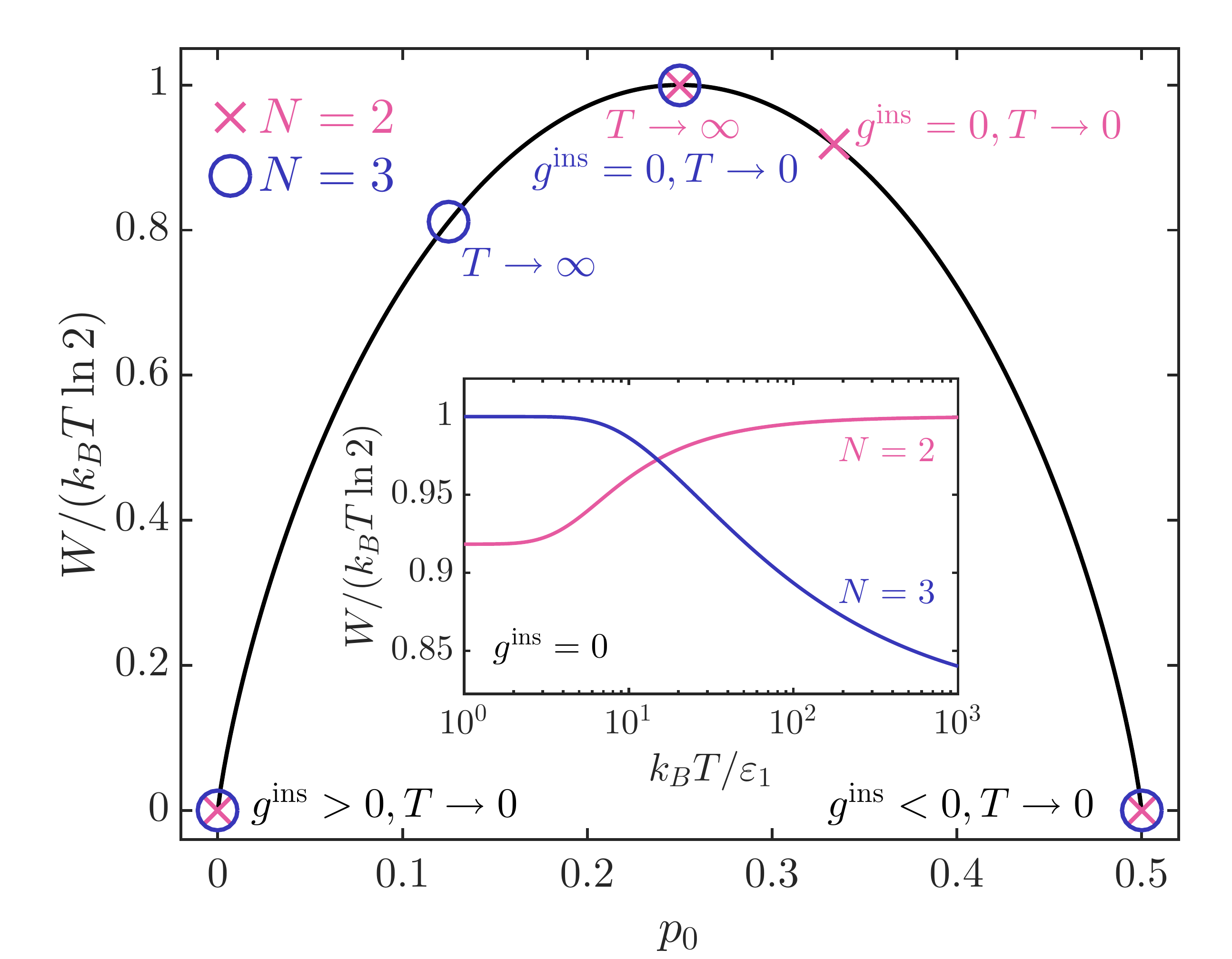}
  \caption{(\textit{Color online}) Work output of the Feshbach-driven Szilard engine
    for $N = 2$ and $N = 3$ as a function of the probability
    $p_0(L/2,g^{\mathrm{ins}})$ to find zero particles to the left of the barrier
    after insertion. {\textit{Inset:}} Work output $W/(k_BT\ln 2)$ for $N = 2$ and $N
    = 3$ as a function of temperature when $g^{\mathrm {ins}}=0$. 
    $\varepsilon_1 = \hbar^2 \pi^2/(2mL^2)$ is the single-particle ground state energy
    in the absence of a barrier.
  }
  \label{fig:feshtt}
\end{figure}

Let us finally investigate the cases of $N = 2$ and $N = 3$, where the attractive
interaction does not seem to increase the maxima in $W/(k_BT\ln2)$.
With a central barrier insertion position,
$p_n(L/2,g^{\mathrm{ins}})=p_{N-n}(L/2,g^{\mathrm{ins}})$ for $n=0,\dots,N$. This
symmetry in $p_n(L/2,g^{\mathrm{ins}})$, combined with the fact that $\sum _n
p_n(L/2,g^{\mathrm{ins}}) =1$, allows us to  express the optimal average work output
per cycle as a function of $p_0(L/2, g^{\mathrm{ins}})$ alone,
\begin{align}
  W  =& -k_B T\Big\{2p_0(L/2, g^{\mathrm{ins}}) \ln\left[2p_0(L/2,
    g^{\mathrm{ins}})\right] \nonumber
  \\ &+ \left[1-2p_0(L/2, g^{\mathrm{ins}})\right]\ln\left[1-2p_0(L/2,
    g^{\mathrm{ins}})\right]\Big\},
  \label{Eq:WorkN2N3}
\end{align}
where the probabilities in
Eqs.~(\ref{Eq:probabilities0},\ref{Eq:probabilities1},\ref{Eq:probabilities2}) are
used to maximize $W$. The work output in Eq.~(\ref{Eq:WorkN2N3}) has a peak value of
$W=k_B T \ln 2 $, which is obtained for $p_0(L/2, g^{\mathrm{ins}}) = 1/4$, see
Fig.~\ref{fig:feshtt}. For $N=3$, the largest ratio $W/(k_BT\ln2)$ is consequently
retrieved with a uniform probability distribution $p_n(L/2,g^{\mathrm{ins}})$, i.e.
with a maximal Shannon information. Such a probability distribution is, in turn,
obtained in the $T\to 0$ limit when $g^{\mathrm{ins}} = 0$. For $g^{\mathrm{ins}} <
0$, the same optimal probability distribution, and thus maximum in $W/(k_BT\ln2)$,
is instead seen at a finite temperature. If we start in the low-temperature limit
(where $p_0(L/2, g^{\mathrm{ins}}) = 1/2$) and continuously increase $T$, then we
will eventually pass through the optimal value $p_0(L/2, g^{\mathrm{ins}}) = 1/4$
towards $p_0(L/2, g^{\mathrm{ins}}) = 1/8$ in the classical limit ($T\to\infty$).
For $N=2$, the maximum in $W/(k_BT\ln2)$, is obtained for $p_1(L/2,
g^{\mathrm{ins}}) = 2p_0(L/2, g^{\mathrm{ins}})$, and thus not at the maximal
Shannon information. Regardless of the (finite) interaction strength, for two
bosons, we approach the optimal value $p_0(L/2, g^{\mathrm{ins}}) = 1/4$ in the
classical limit.

\section{Feshbach-Assisted Szilard Engine}

Let us now incorporate the interaction-tuning step into the conventional cycle of
the many-particle quantum Szilard engine~\cite{kim2011,kim2012,bengtsson2018}. In
other words, in addition to choosing $g_n^{\mathrm{rem}}$ optimally to maximize the
work output  we now also simultaneously choose the optimal values for the removal
positions $\ell_n^{\mathrm{rem}}$. We consider a central insertion position, i.e.
$\ell^{\mathrm{ins}} = L/2$, which was found optimal in the region of largest
$W/(k_BT\ln 2)$
for non-interacting and attractively interacting bosons in the conventional Szilard
engine~\cite{bengtsson2018}.
\begin{figure}
  \centering
  \includegraphics[width = 0.5\textwidth]{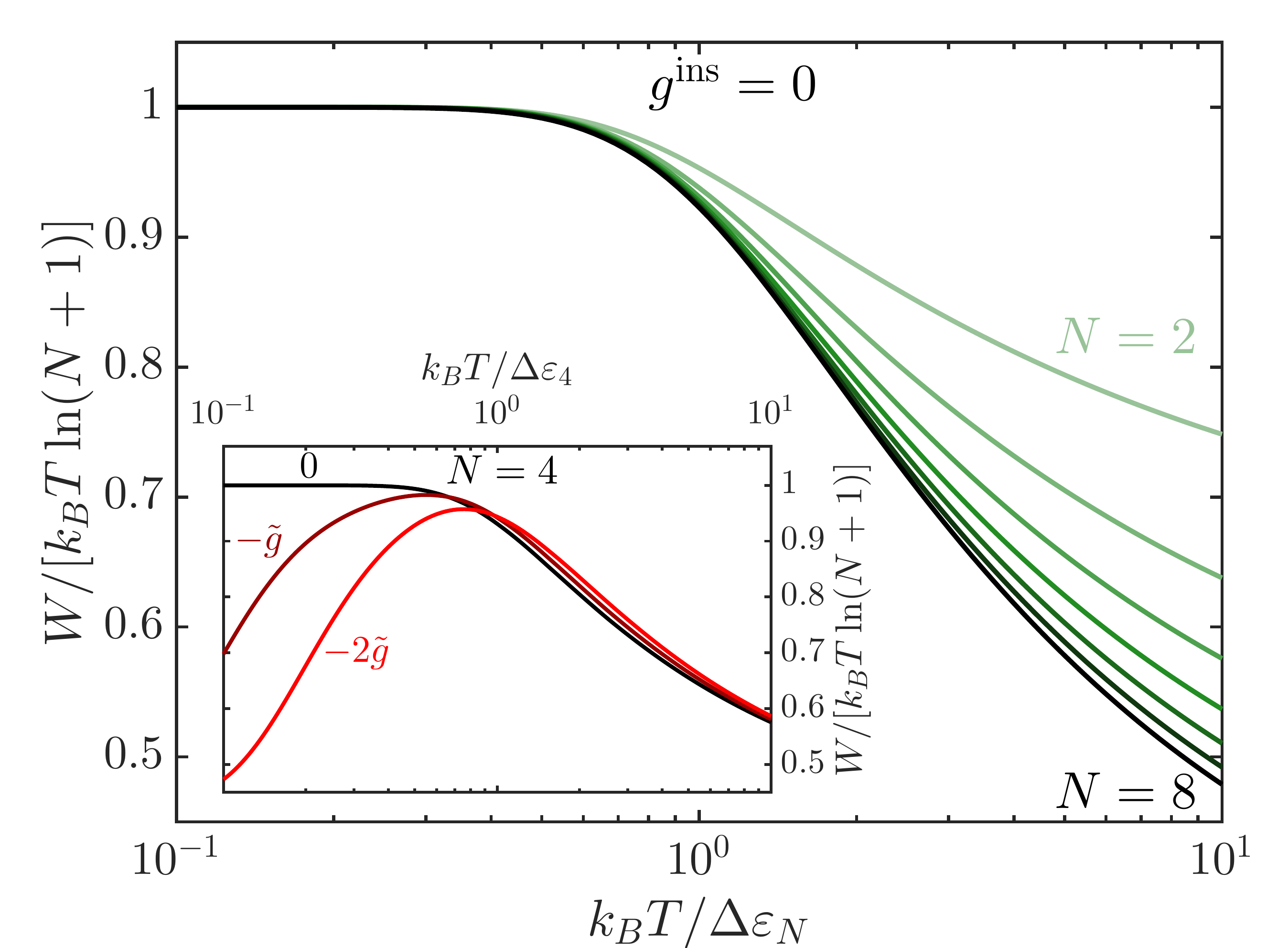}
  \caption{(\textit{Color online}) Relative work output, $W/[k_BT\ln (N+1)]$, of the
    Feshbach-assisted Szilard engine as a function of temperature $T$ for $N \le 8$
    particles, with $\Delta \varepsilon_N = \epsilon_{N+1} - \epsilon_N$. The
    different lines  are for different particle numbers, from $N=2$ (top) to  $N=8$ (bottom). 
    {\textit {Inset:}} Effect of different interaction strengths $g^{\mathrm{ins}}$ at insertion (as indicated in the figure) for the
    example of $N=4$ particles in the engine. The relative work output is
    significantly lowered in the deep quantum regime, but increases at higher
    temperature for an initial attraction.}
  \label{fig:Tdep}
\end{figure}
The work output of the Feshbach-assisted Szilard engine with $N\le 8$ bosons is
shown in Fig.~\ref{fig:Tdep}. Note that we here show the ratio $W/[k_BT\ln(N+1)]$ as
a function of $T$. Of key interest is  that the {\it maximal} possible average
work output, only bounded by the second law for feedback processes
\cite{parrondo2015}, can be achieved at low temperatures for $g^{\mathrm{ins}}=0$.
First, we recall that the Shannon information is maximal for bosons in the $T\to 0$
limit when $\ell^{\mathrm{ins}} = L/2$ and $g^{\mathrm{ins}} = 0$. Secondly, the
full information-to-work conversion may be explained by the fact that in a
one-dimensional system, bosons with infinitely strong repulsive contact-interactions
act as spin-polarized fermions, i.e., a Tonks-Girardeau~\cite{Girardeau1960} gas. In other words, when
$g^{\mathrm{rem}}_n \to \infty$ for all $n=0,\ldots,N$,  the fermionized bosons fill
up the single-particle energy levels according to the Pauli exclusion principle. A
barrier position may then always be found such that the many-particle ground state
consists of $n$ particles to the left of the barrier  and $N-n$ to the right. In the
considered low-temperature limit, the removal process can thus always be made fully
reversible with a complete information-to-work conversion. A straightforward
calculation reveals the intervals for optimal removal,
\begin{equation}
  \frac{n}{N + 1} < \frac{\ell_n^{\mathrm{rem}}}{L} < \frac{n + 1}{N + 1}.
  \label{Eq:remopt}
\end{equation}
Also with $g^{\mathrm{ins}}<0$, a maximal information-to-work conversion efficiency
is possible in the low-temperature limit. However, since all the particles
necessarily are found on the same side of the barrier upon measurement, the Shannon
information is drastically reduced and, as a consequence, also the work output, see
the inset of Fig.~\ref{fig:Tdep}.

If we increase the temperature, the work output can be seen to decrease when
$g^{\mathrm{ins}}=0$ (see Fig.~\ref{fig:Tdep}). This degradation of the engine is of
twofold origin: Partly, it occurs due to a reduction in the Shannon information, and
partly it is because of a loss of reversibility associated with the barrier removal
process. With increased temperature, the distribution of measurement outcomes goes
from $p_n(L/2, 0) \rightarrow 1/(N+1)$ (in the $T\to 0$ limit) to the classical
distribution where the particles behave as distinguishable ones, $p_n(L/2, 0)
\rightarrow \binom{N}{n}/2^N$ (in the $T \to \infty$ limit). The latter distribution
is peaked about $n\sim N/2$ and thus has a lower Shannon information. In the case of
$g^{\mathrm{ins}}<0$, on the other hand, the information will typically grow with
$T$ initially, when more measurement outcomes becomes accessible, but will later
decay as the classical limit is approached. Similar to the work output of the
Feshbach-driven engine seen in Fig.~\ref{fig:wtdep}, the ratio $W/[k_BT\ln(N+1)]$
now has a maximum (although typically smaller than the maxima for
$g^{\mathrm{ins}}=0$) at a finite temperature (see the inset of
Fig.~\ref{fig:Tdep}). Furthermore, also similar to the Feshbach-driven engine, it is
important to also account for the information-to-work conversion losses when
establishing the details of this maximum.

\begin{figure}
  \centering
  \includegraphics[width = 0.5\textwidth]{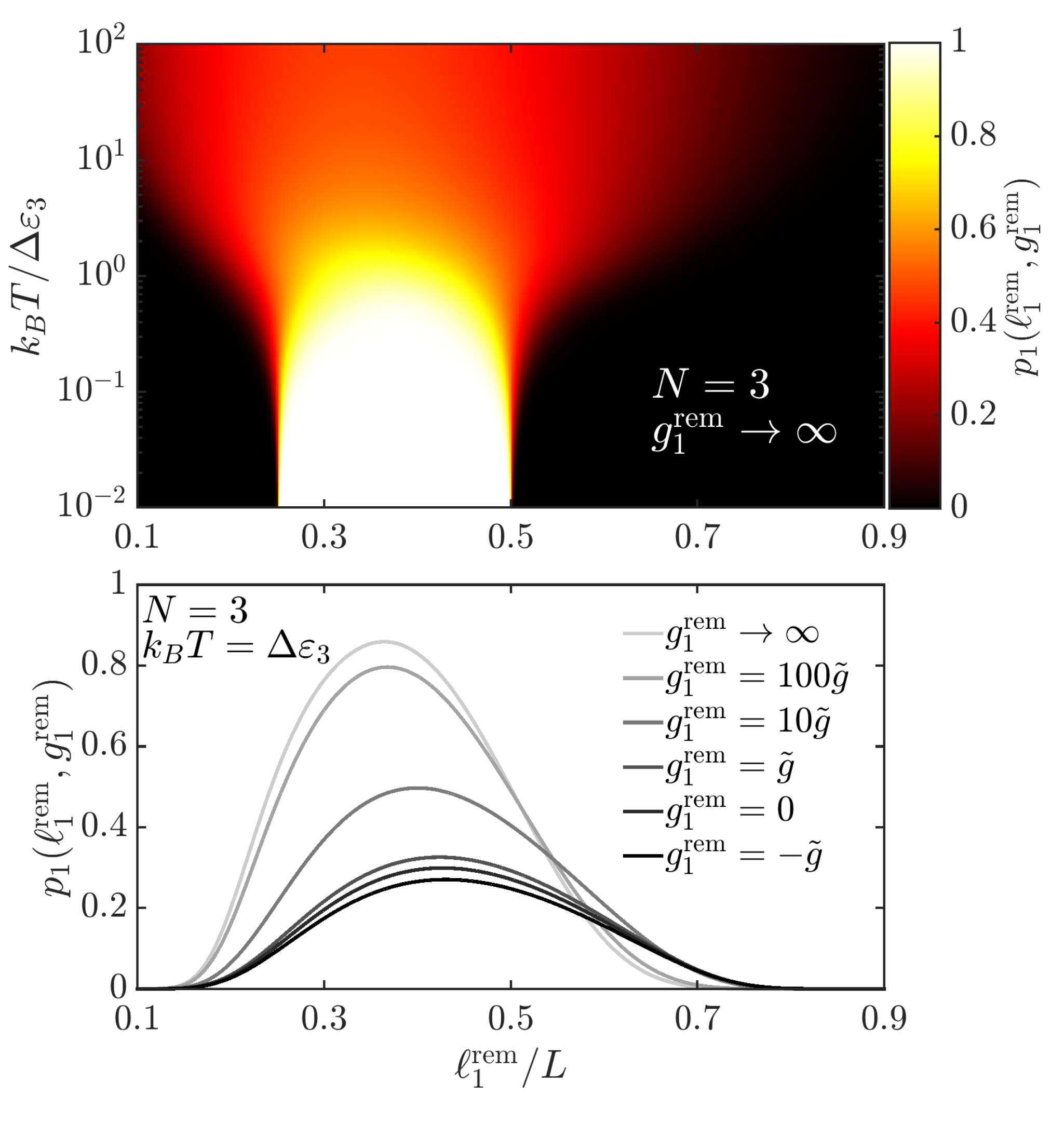}
  \caption{(\textit{Color online}) \textit{Upper panel:} The reversibility
    probability $p_1(\ell^{\mathrm{rem}}_1,g_1^{\mathrm{rem}})$ as a function of the
    removal position $\ell^{\mathrm{rem}}_1$ and temperature $T$ at
    $g_1^{\mathrm{rem}}\rightarrow \infty$, i.e., at complete fermionization, for the
    case of $N=3$ particles. \textit{Lower panel:}
    $p_1(\ell^{\mathrm{rem}}_1,g^{\mathrm{rem}}_1)$ as a function of
    $\ell^{\mathrm{rem}}_1$ at $k_B T/\Delta \varepsilon_3 = 1$ for different values
    of $g_1^{\mathrm{rem}}$. We can see that optimal removal still occurs at complete
    fermionization even though the removal process is no longer reversible.}
  \label{fig:rev}
\end{figure}

The second reason for the reduction in $W/[k_BT\ln(N+1)]$, namely the growing losses
in the information-to-work conversion with $T$, is illustrated for the case of $N=3$
in Fig.~\ref{fig:rev}. In particular, we consider the case in which one boson is
found to the left of the barrier and show the probability
$p_1(\ell^{\mathrm{rem}}_1,g^{\mathrm{rem}}_1\to \infty)$ as a function of the
barrier removal position. Clearly, in the limit of $T\to 0$, a fully reversible
removal process is possible, in agreement with the condition in
Eq.~(\ref{Eq:remopt}). As the temperature increases, excited many-body states starts
to become more populated during the barrier removal process. The probability
$p_1(\ell^{\mathrm{rem}}_1,g^{\mathrm{rem}}_1\to \infty)$ will then decrease,
indicating growing losses in the information-to-work conversion. We may finally
estimate the onset of these conversion losses based on that the thermal energy has
to be of the order of $k_B T \sim  \Delta \varepsilon _N=\varepsilon
_{N+1}-\varepsilon_N = (N+1)^2\varepsilon _1 - N^2\varepsilon _1 = (2N +
1)\varepsilon _1$ for any excited many-body state to have a significant population.
Note that the temperatures shown in Fig.~\ref{fig:rev}, as well as in
Fig.~\ref{fig:Tdep}, are  scaled by the factor $\Delta \epsilon_N$.

\section{Conclusion}

This work suggests an optimal protocol for the quantum Szilard engine, where the interaction
strength of the working medium is allowed to vary according to the measurement
outcome. We have shown that by adding this new aspect to the regular scheme of the
Szilard engine, it can be made fully reversible in the quantum regime and reach the
maximal possible work output per cycle as dictated by the second law of thermodynamics. For a sufficiently strong
repulsion between the quantum particles constituting the engine's medium and at low enough temperatures, there exists an interval of values for each
removal position such that the engine is made reversible. Furthermore,
we have seen that deterioration in work output at higher temperatures can be
decreased by improving the engine's information content through the addition of an
initial attraction to the working medium.
The setup suggested here could be realizable for example with ultra-cold atoms where the interactions can be controlled by Feshbach resonances.
Our work opens important new perspectives for information-driven quantum heat engines, for which the concept of the Szilard cycle since long has been the prime paradigm. 

\begin{acknowledgements}
  We thank G.~Kir\v{s}anskas for very valuable comments and his help
  regarding the implementation of the Bethe ansatz.  We also thank T.~Busch,
  H.~Linke, P.~Samuelsson,  M.~Ueda and A.~Wacker  for many helpful discussions on
  the many-body Szilard engine. This work was financially supported by Knut and
  Alice Wallenberg Foundation,
  NanoLund and the Swedish Research Council.

\end{acknowledgements}

\bigskip

\appendix*
\section{Derivation of Work Expression}\label{appa}

For the derivation for the work expression for the Feshbach-assisted Szilard engine, we follow very closely the concept suggested by Kim {\it et al.}~\cite{kim2011} that was also applied in Ref.~\cite{bengtsson2018}. 
The average work output associated with a change of the length and interaction strength parameters 
from some initial values $(\ell _i, g_i)$  to some final values $(\ell _f, g_f)$  in an isothermal process is given by
\begin{equation*}
  W_{\mathrm{iso}} = k_B T \ln \left[\frac{Z(\ell _f, g_f)}{Z(\ell _i, g_i)}\right]
\end{equation*}
where $Z = \sum_n e^{-E_n/(k_BT)}$ is the partition function. In the first step of the cycle, (i), a barrier is raised at $\ell^{\mathrm{ins}}$
for $N$ particles with interaction strength $g^\mathrm{ins}$, which costs an amount
of work equal to
\begin{equation*}
  W_{(i)} = k_B T\ln
  \left[\frac{\sum_{n=0}^NZ_n(\ell^{\mathrm{ins}},g^{\mathrm{ins}})}{Z_N(L,
      g^{\mathrm{ins}})}\right].
\end{equation*}
Here $Z_n(\ell, g)$ is the partition function obtained when the sum runs over the energies with $n$ particles to the left of the barrier alone.
In the second step, the
number of particles on each side is measured and subsequently an expansion is performed
until the barrier reaches $\ell^{\mathrm{rem}}_n$.
The work associated with this expansion/compression step of the cycle is given by
\begin{equation*}
  W_{(ii)}^{(1)} = k_B T\sum_{n=0}^N p_n(\ell^{\mathrm{ins}},g^{\mathrm{ins}}) \ln
  \left[\frac{Z_n(\ell^{\mathrm{rem}}_n,
      g^{\mathrm{ins}})}{Z_n(\ell^{\mathrm{ins}},g^{\mathrm{ins}})}\right],
\end{equation*}
where $p_n = Z_n/\sum_{m=0}^N Z_m$ is the 
probability of measuring n particles to the left of the barrier.

Furthermore, the interaction strength is changed to
$g_n^\mathrm{rem}$ depending on the measurement outcome, giving the additional work
\begin{equation*}
  W_{(ii)}^{(2)} = k_B T\sum_{n=0}^N p_n(\ell^{\mathrm{ins}},g^{\mathrm{ins}}) \ln
  \left[\frac{Z_n(\ell^{\mathrm{rem}}_n,
      g_n^{\mathrm{rem}})}{Z_n(\ell^{\mathrm{rem}}_n,g^{\mathrm{ins}})}\right].
\end{equation*}
Next, in step (iii), the barrier is removed, and as its height is lowered, particles
will eventually be able to tunnel between the two systems. Provided that the processes are
carried out quasi-statically, the average work associated with this process reads (see Ref.~\cite{kim2011})
\begin{equation*}
  W_{(iii)} = k_B T\sum_{n=0}^N p_n(\ell^{\mathrm{ins}},g^{\mathrm{ins}}) \ln
  \left[\frac{Z_N(L, g_n^{\mathrm{rem}})} {\sum_{n=0}^N
      Z_n(\ell^{\mathrm{rem}}_n,g_n^{\mathrm{rem}})}\right].
\end{equation*}
In the fourth and last step, the interaction strength is changed back to its
original value, from $g_n^\mathrm{rem}$ to $g^\mathrm{ins}$, and  
\begin{equation*}
  W_{(iv)} = k_B T\sum_{n=0}^N p_n(\ell^{\mathrm{ins}},g^{\mathrm{ins}}) \ln
  \left[\frac{Z_N(L, g^{\mathrm{ins}})}{Z_N(L, g_n^{\mathrm{rem}})}\right].
\end{equation*}
Finally we get the total average work output per cycle by summing up the different
contributions $W = W_{(i)}+W_{(ii)}^{(1)}+W_{(ii)}^{(2)}+W_{(iii)}+W_{(iv)}$.


%

\end{document}